\documentclass{emulateapj}
\usepackage{apjfonts}
\usepackage{ulem}
\usepackage{color}

\def\lsim{\;\raise0.3ex\hbox{$<$\kern-0.75em\raise-1.1ex\hbox{$\sim$}}\;}
\def\gsim{\;\raise0.3ex\hbox{$>$\kern-0.75em\raise-1.1ex\hbox{$\sim$}}\;}
\def \cmsec {\rm ~cm~s^{-1}}
\shorttitle{}
\shortauthors{}
\newcommand{\Lmin}{L_\mathrm{min}}
\newcommand{\Lmax}{L_\mathrm{max}}
\newcommand{\Kmin}{k_\mathrm{min}}
\newcommand{\Kmax}{k_\mathrm{max}}
\newcommand{\Vph}{v_\mathrm{ph}}
\def\kms{\rm ~km~s^{-1}}

\def\cmc{\rm ~cm^{-3}}

\begin{document}

\title{Dots, clumps and filaments: the intermittent  
images of synchrotron emission in random magnetic fields
of young supernova remnants}

\author{Andrei M. Bykov\altaffilmark{1},  Yury A. Uvarov\altaffilmark{1} and
Donald C. Ellison\altaffilmark{2} }


\altaffiltext{1}{Ioffe Institute for Physics and Technology, 194021 St.Petersburg,
 Russia; byk@astro.ioffe.ru}
\altaffiltext{2}{Physics Department, North Carolina State
University, Box 8202, Raleigh, NC 27695; don\_ellison@ncsu.edu}

\begin{abstract}Non-thermal X-ray emission in some supernova remnants
originates from synchrotron radiation of ultra-relativistic
particles in turbulent magnetic fields. We address the effect of a
random magnetic field on synchrotron emission images and spectra.
A random magnetic field is simulated to construct synchrotron
emission maps of a source with a steady distribution of
ultra-relativistic electrons. Non-steady localized structures
(dots, clumps and filaments), in which the magnetic field reaches
exceptionally high values, typically arise in the random field
sample. These magnetic field concentrations dominate the
synchrotron emission (integrated along the line of sight) from the
highest energy electrons in the cut-off regime of the
distribution, resulting in an evolving, intermittent, clumpy
appearance. The simulated structures resemble those observed in
X-ray images of some young supernova remnants. The lifetime of
X-ray clumps can be short enough to be consistent with that
observed even in the case of a steady particle distribution. The
efficiency of synchrotron radiation from the cut-off regime in the
electron spectrum is strongly enhanced in a turbulent field
compared to emission from a uniform field of the same magnitude.
\end{abstract}


\section{Introduction}
The forward shocks of supernova remnants (SNRs) are the most
probable accelerators of cosmic rays to high energies.
Supporting this idea is the believe that the diffusive shock
acceleration (DSA) mechanism is capable of accelerating particles
to above 100 TeV in young and middle age SNRs.
Electrons (and/or positrons) accelerated to $\sim \!10$ TeV will
efficiently radiate in X-rays if magnetic fields exceed 100 $\mu$G.
With {\sl Chandra's} imaging capability it is possible to
distinguish the synchrotron structures in the X-ray images of Cas A,
RCW 86, Kepler, Tycho and some other SNRs
\citep[e.g.,][]{vl03,bambaea06,pf08}. The non-thermal emission seen
by {\sl Chandra} concentrates in very thin filaments in all SNRs.
\citet{bambaea06} also reported on the time evolution of the scale
length of the filaments and roll-off frequency of the SNR spectra.

The synchrotron filamentary structures could be due to a narrow
spatial extension of the highest energy electron population in the
shock acceleration region limited by efficient synchrotron energy
losses.
In this case, strong magnetic field amplification in the shock
vicinity is required to match the observed thickness of the
non-thermal filaments \citep[e.g.,][]{vl03,bambaea05, vbk05}, thus
supporting the case for highly efficient DSA. An alternative
interpretation, that the observed narrow filaments are limited by
magnetic field damping and not by the energy losses of the radiating
electrons, was proposed by \citet{pyl05}.

There is an emerging class of SNRs (SN1006, RXJ1713.72-3946, Vela
Jr, etc) dominated by non-thermal X-ray emission. These SNRs are of
particular interest since their X-ray emission most likely
originates from synchrotron radiation of electrons accelerated to
well above TeV energies.
Many of the non-thermal X-ray SNRs are known to be radio-faint compared
with most SNRs, and some of these remnants were first discovered in
X-rays.
The sources mentioned are now known to emit TeV photons as well and the
origin of this emission is of fundamental importance for the origin of
cosmic rays.
Is it inverse-Compton emission from the X-ray emitting electrons, or
pion-decay emission from proton-proton interactions, or both?
%
One way to get the
answer is to study the magnetic field structure in the SNRs.

Recently \citet{uchiyamaea07} reported variability of X-ray hot
spots in the shell of SNR RXJ1713.72-3946 on $\sim$ one-year
timescales. The authors suggested that this X-ray variability results
from radiation losses and, therefore, reveals
the ongoing shock-acceleration of electrons in real time.
In order for radiation losses to be this rapid, the underlying magnetic
field would need to be amplified by a factor of more than 100.
From multi-wavelength data analysis \citet{buttea08} concluded
however, that the mG-scale magnetic fields estimated by
\citet{uchiyamaea07} cannot span the whole non-thermal SNR shell,
and if small regions of enhanced magnetic field do exist in RX
J1713-3946, it is likely that they are embedded in a much weaker
extended field. Therefore, it is essential to study the effects of
magnetic field structure and particle distributions on the observed
synchrotron emission maps.

Turbulent magnetic fields, with energy densities approaching a
substantial fraction of the shock ram pressure, are a generic
ingredient in the efficient DSA mechanism. However, the effect of
turbulence on synchrotron emission images and spectra has not yet
been addressed. In this paper, we model specific features in
synchrotron images arising from the stochastic nature of magnetic
fields. In \S{\ref{fields}} we describe our random magnetic field
simulations, and these are  used in \S{\ref{synchr}} to construct
synchrotron emission maps for different frequencies.
Our results show that the rapid time variations seen in some SNRs
may well be produced by {\sl steady} electron distributions in
turbulent fields.

\section{Simulation of a random magnetic field}\label{fields}
In order to simulate numerically an isotropic and spatially
homogeneous turbulent field we sum over a large number ($N_m$) of
plane waves with wave vector, polarization, and phase chosen
randomly \citep[see e.g.][]{gj99,casseea02,cr04}, i.e.,
\begin{equation}
{\bf B}({\bf r}, t)=\sum_{n=1}^{N_m}\sum_{\alpha=1}^2{\bf
A}^{(\alpha)}(k_n) \cos({\bf k}_n\cdot{\bf r} -\omega_n({\bf
k}_n)\cdot t + \phi_n^{(\alpha)})\ , \label{bf1}
\end{equation}
where the two orthogonal polarizations ${\bf A}^{(\alpha)}(k_n)$
($\alpha=1,2$) are in the plane perpendicular to the wave vector
direction (i.e., ${\bf A}^{(\alpha)}(k_n)\perp{\bf k}_n$, so as to
ensure that $\nabla\cdot{\bf B}=0$). We divided k-space between
$\Kmin=2\pi/\Lmax$ and $\Kmax=2\pi/\Lmin$ into $N$ spherical shells
distributed uniformly on a logarithmic scale (a number of points in
the s-shell is $M_s$). Here $\Lmin$ and $\Lmax$ are the minimum and
maximum scales of turbulence respectively.
  The spectral energy
density of the magnetic field fluctuations is taken as $W(k) \propto
k^{-\delta}$, where  $\delta$ is the spectral index
\citep[c.f.][]{gj99}. The average square magnetic field $\langle
B^2\rangle = \int dk\, W(k)$ is an input parameter for our model.
In the particular simulation runs presented
below we assume $\omega_n({\bf k}_n) = v_{\rm ph}\cdot k_n$. The
parameter $v_{\rm ph}$ would be a phase velocity in the case of a
superposition of linear modes (e.g., magnetosonic waves). In this
paper, however, we are modeling strong magnetic field fluctuations,
so the random Fourier harmonics in Eq.(\ref{bf1}) comprising the
random field should not be exactly associated with any MHD eigen
modes.

Following the prescription given above in Eq.(\ref{bf1}), we
simulated the temporal evolution of random magnetic fields filling a
cube of scale size $D$, where $\Lmin= 2\times 10^{-4} \pi D$,
$\Lmax=0.2\, \pi D $, $N_m=4800$, and $N=6$, $M_s = 800$. The choice
of the mode number $N_m=4800$ provided the simulated fluctuation
spectrum to be fitted with power law of index $\delta= 1.0$ between
$\Kmin$ and $\Kmax$ with accuracy better than 2$\%$. The spectral
index $\delta =1$ corresponds to the so-called Bohm diffusion model
in DSA.
With the simulated random magnetic fields, we calculated the local
spectral emissivity of polarized synchrotron emission at various times
and for different distributions of electrons or positrons. The technique
is described next along with the simulated emission maps.


\begin{figure} \epsscale{1.2} \plotone{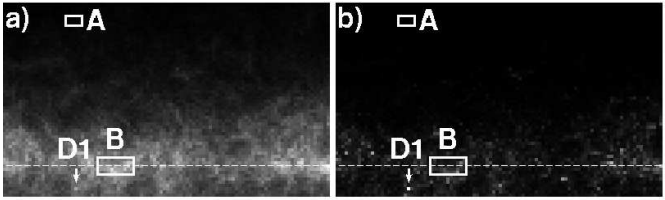}\caption{Panels (a) and (b) are simulated maps
of synchrotron emission in a random magnetic field. Intensities,
$\nu^2\cdot I({\bf R_{\perp}},t, {\nu})$, are shown with a linear
grayscale (white is maximum). The map (a) is made at 5 keV, while
(b) is made at 50 keV.  The shock position, where the electron
  distribution peaks, is indicated with a dashed line. The stochastic magnetic field
  sample has $\sqrt{\langle B^2\rangle}
= 10^{-4}$ G and $\delta = 1.0$. A dim region ``A'' is marked to
contrast a bright region "B" and a dot marked as ``D1''.}
\label{syn_map}
\end{figure}

\begin{figure*}[t]
\includegraphics[width=0.99\textwidth]{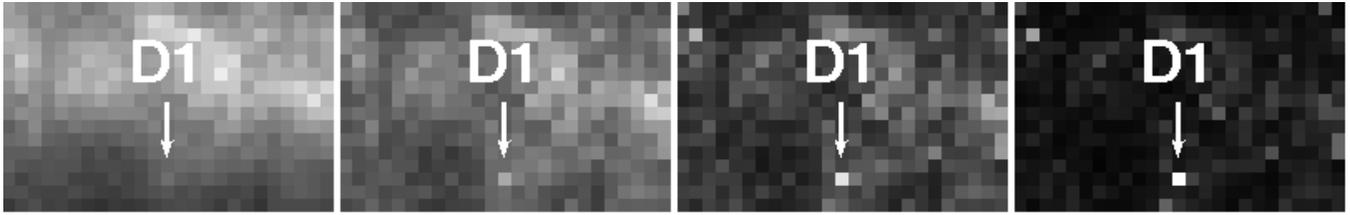}
\caption{Details of Fig.\ref{syn_map} in the vicinity of the bright
dot D1. The four panels show, from left to right, $\nu^2\cdot I({\bf
R_{\perp}},t, {\nu})$ maps of synchrotron emission at 0.5, 5, 20 and
50 keV, respectively.} \label{4nu_map}
\end{figure*}


\section{Synchrotron image simulations}\label{synchr}
 The formation length $l_f$ of the synchrotron radiation pattern of an
individual ultra-relativistic electron is energy independent, i.e.,
$l_f \approx R_g/\gamma = mc^2/eB \approx 1.8\times 10^9~B_{\mu
G}^{-1}$ cm, where $R_g$ is the gyro-radius of a particle with
Lorentz factor $\gamma$. The length $l_f$ is a small ($\gamma^{-1}$)
fraction of the particle gyro-radius.
In this study we only consider the effects of magnetic fluctuations
having scales comparable to the gyro-radii of ultra-relativistic
particles that are much larger than $l_f$.
Therefore, the fluctuation wavenumbers $k$ satisfy $k \cdot l_f \ll
$ 1.
This restriction allows us to apply the standard formulae for the
synchrotron power of a single particle of Lorentz factor $\gamma \gg
1$ in a homogeneous magnetic field, and then to integrate this power
over the line of sight through the system filled with random field
fluctuations.

We start with the spectral flux densities $p^{(1)}_{\nu}(\theta,E)$
and $p^{(2)}_{\nu}(\theta,E)$ with two principal directions of
polarization radiated by a particle with energy $E$, given by
\citet{gs65} [their Eqs.(2.20)]. Here $\theta$ is the angle between
the local magnetic field ${\bf B}({\bf r}, t)$ and the direction to
the observer. In the case of a random magnetic field it is
convenient to use the local spectral densities of the Stokes
parameters \citep{gs65}. A synchrotron map of a source can be
constructed as a function of the position ${\bf R_{\perp}}$ in the
plane perpendicular to the observer direction from the photons of a
fixed frequency. Because of the additive property of the Stokes
parameters
 for incoherent photons,
we can integrate $p^{(1)}_{\nu} + p^{(2)}_{\nu}$ over the line of
sight in the observer direction weighted with the distribution
function of emitting particles $N({\bf r}, E, t)$ to get intensity
$I({\bf R_{\perp}}, t,{\nu})$.
To collect the photons reaching the source surface at the same
moment, $t$, we
performed the integration 
over the source depth using the retarded time $t' = t - |{\bf
r}-{\bf R_{\perp}}|/c$ as arguments in ${\bf B}({\bf r}, t')$ and
$N({\bf r}, E, t')$. The integration grid had cell size smaller than
$\Lmin$. We produce the maps at different frequencies $\nu$ and for
different times to study spectra and variability. Below we present
synchrotron images simulated with a steady model distribution of
electrons accelerated by a plane shock of velocity 2,000 $\kms$
propagated in a fully ionized plasma of number density 0.1$\cmc$.
The kinetic model used to simulate the electron distribution was
described in detail by \citet{bceu00}. The magnetic field in the
far-upstream region was fixed at 10
  $\mu$G and it was assumed that magnetic field amplification produced a
  random field of $\sqrt{\langle B^2\rangle} = 10^{-4}$ G in the shock
  vicinity. The effective shock compression ratio is about 4.7.

The spatial distribution of $\sim$ TeV  electrons is highly
non-uniform and concentrated near the shock due to synchrotron
losses \citep[see, also][]{za07}. The simulated electron energy
spectrum at the shock front position has asymptotic behavior $N
\propto 7\times 10^{-12} [E/\mathrm{TeV}]^{-1.8} \exp[-(E/7\,\mathrm{TeV})^2]$
$\cmc\,TeV^{-1}$  in the cut-off energy regime.

We note that, in reality, the distribution of the emitting electrons
would be a random function of the position and particle energy
because of the stochastic nature of both the electromagnetic fields
and the particle dynamics.  However, such a stochastic distribution
can currently only be determined with particle-in-cell (PIC)
simulations and then only for very limited energy ranges. The PIC
simulation of TeV distributions in SNR shocks is beyond current
computer capabilities \citep[see, ][]{vbe08}.

Our model highlights the effects of a stochastic magnetic field on
the synchrotron mapping and demonstrates the importance of the
high-energy cutoff region of the electron distribution where the
synchrotron emissivity depends strongly on the local magnetic field.

\section{Mapping of a synchrotron emission source}\label{maps}
In this section we calculate maps
of a synchrotron emission source with a random magnetic field. We
use a set of parameters for a stochastic magnetic field sample with
$\sqrt{\langle B^2\rangle} = 10^{-4}$ G and $\delta = 1.0$. Note
that a 7 TeV electron has a characteristic synchrotron emission
energy $h \nu_{\rm max} \approx$ 0.1 keV (calculated for a uniform
magnetic field $B = 10^{-4}$\,G). However, this energy is not a good
spectral characteristic in a random magnetic field because of the
presence of the long tails in the probability distribution function
(PDF) of the random field, as is clear in Figure~\ref{sp}. In
Figure~\ref{syn_map} we show $\nu^2\cdot I({\bf R_{\perp}},t,
{\nu})$ maps at some particular time. Shock is propagating
perpendicular to the line of sight direction. The 5 keV image (left
panel a) clearly demonstrates the presence of structures: filaments,
clumps and dots produced by the stochastic field topology. Filaments
are less prominent at 50 keV (panel b), while dots and clumps are
present. Polarization maps, where the degree of polarization is
determined in a standard way also show clumpy structures. The degree
of polarization reaches very high values in high energy clumps. We
will discuss in detail the polarization properties of synchrotron
structures and clump statistics elsewhere.

In Figure~\ref{4nu_map}, the synchrotron emission of a small region
around the dot D1 is shown for 0.5, 5, 20 and 50 keV. There are
observable differences in the synchrotron maps indicating that some
features are bright at high energies and much less prominent at
lower energies and vice versa.
The physical reason for this difference comes from the fact that in
the high-energy cut-off region of the electron spectrum, the local
synchrotron emissivity depends strongly on the local field value.
High order statistical moments of the field dominate the synchrotron
emissivity at high energies, and even a single strong local field
maximum can produce a feature (dot or clump) that stands out on the
map even after integrating the local emissivity over the line of
sight. In lower energy maps, the contribution of a single maximum
can be smoothed or washed out by contributions from a number of
weaker field maxima integrated over the line of sight. The high
energy map is highly intermittent because the synchrotron emissivity
depends on high order moments of the random magnetic field at the
cut-off frequency regime.

\subsection{Spectra and temporal evolution of structures}\label{spec}
\begin{figure}
\epsscale{1.2} \plotone{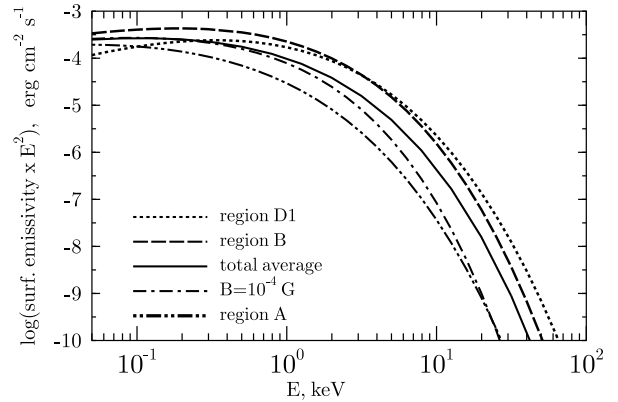} \caption{Spectral densities of
surface emissivity for regions in Fig.~\ref{syn_map} as labeled. The
solid curve is the average over the entire region shown in
Fig.~\ref{syn_map}, while the dot-dashed curve is the spectrum
produced if the region is filled with a uniform magnetic field of
fixed magnitude $B$=10$^{-4}$ G, but of random direction.}
\label{sp}
\end{figure}

In Figure~\ref{sp} we show the simulated spectral density of the
surface emissivity of a synchrotron source of size $D$ = 0.3 pc with
the magnetic field model described above.
The spectra for the regions of different surface emissivity apparent
in Figure~\ref{syn_map} are plotted separately in Figure~\ref{sp}.
A dim rectangular region A and a bright dot D1 indicated in
Figure~\ref{syn_map} have drastically different spectra above 0.1
keV in the model. At the same time the spectrum of D1 is very
similar to the total emissivity averaged over the whole map,
indicating that the total emissivity is dominated by the bright dots
and clumps.

The fact that bright dots and clumps contribute so strongly to the
total emission means that there will be a clear spectral difference
between the case where a source is filled with a uniform magnetic
field (i.e., one equal to 0.1\, mG in our simulation), and the case
where the field is turbulent with $\sqrt{\langle B^2\rangle}=
0.1$~mG, as we simulate in Fig.\ref{syn_map}. The presence of
magnetic field fluctuations can substantially increase the
synchrotron emissivity of electrons in the exponential tail of the
distribution and if this effect is neglected, average magnetic
fields may be inferred that are factors of several larger than is in
fact the case.

\begin{figure}
\epsscale{1.25} \plotone{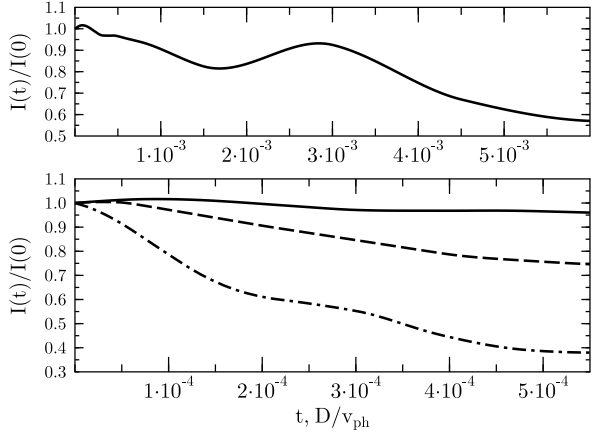} \caption{Temporal evolution of dot
D1 shown in Fig.\ref{4nu_map} at 50 keV (dot-dashed curve), 20 keV
  (dashed curve), and 5 keV (solid line). The upper panel shows the evolution of D1 at 5 keV on a longer time
scale.}
  \label{time}
\end{figure}

An important aspect of synchrotron emission in turbulent fields is
that the emission will vary in time even for a static electron
distribution. In Figure~\ref{time} we show the relative intensity of
the bright unresolved dot D1 [of size $\sim 10^{-2}D$] as a function
of time measured in units of $D/\Vph$, where
$\Vph$ is a characteristic speed of the magnetic turbulence, e.g.,
the Alfv\`en speed. Figure~\ref{time} shows that a factor of two or
more change in intensity will occur on times of a year if $D/v_{\rm
ph} \sim 10^{10}$ s for X-rays with energies $\gtrsim 5$\,keV. We
simulate the image with minimal pixel size $10^{-2}D =10^{16}$ cm,
while $\Lmin$ was $6\times 10^{14}$ cm. For SNRs at kpc distances
the simulated pixel size corresponds to the $\sim arcsecond$ range
resolution of {\sl Chandra}. The minimal variability time scale
depends on the size of the structure $l$ if $l > \Lmin$ and the
dependence on $l$ is nonlinear since the photon emission is
determined by high statistical moments of the random field in the
cutoff regime. A one year
  variation-time of D1 at 5 keV requires $v_{\rm ph} \lesssim
10^{8}\cmsec$, which is a realistic value for the Alfv\`en
  velocity for SNRs if
the magnetic field is $\gsim 0.1$\,mG.
Figure~\ref{time} shows that higher energy X-rays in bright spots
can vary on much shorter time scales, another consequence of
emission from the exponential tail of the electron distribution.
Some of the large scale filamentary structures seen in
Fig.\ref{syn_map} also show a complex time variability.

\section{Conclusions}
We show that prominent, evolving, localized structures (e.g., dots,
clumps and filaments) can appear in synchrotron maps of extended sources
with random magnetic fields, even if the particle distribution is smooth
and steady.
The bright structures originate as high-energy electrons radiate
efficiently in local enhancements of the magnetic field intensity.
The peaks in synchrotron emission maps occur due to high-order
moments of the magnetic field probability distribution function
(PDF).

Even if the PDF of projections of the local magnetic field is nearly
Gaussian, the corresponding PDF of synchrotron peaks, simulated for
a spatially homogeneous relativistic particle distribution, has strong
departures from the Gaussian at large intensity amplitudes.
This is because of the non-linear dependence of the synchrotron
emissivity on the local magnetic field in the high-energy cut-off regime
of the electron spectrum.

Our model makes two basic predictions. One is that the overall
efficiency of synchrotron radiation from the cut-off regime in the
electron spectrum can be strongly enhanced in a turbulent field with
some $\sqrt{\langle B^2\rangle}$, compared to emission from a uniform
field, $B_0$, where $B_0 = \sqrt{\langle B^2\rangle}$.
The second is that strong variations in the brightness of small
structures can occur on time scales much shorter than variations in
the underlying particle distribution. The variability time scale is
shorter for higher energy synchrotron images.
Our estimate of the time scales of these intensity variations is
consistent with the rapid time variability seen in some young SNRs
by {\sl Chandra}. The strong energy dependence we predict may be
important for the future missions {\sl NuSTAR}, {\sl Simbol-X} and
{\sl GRI} that will image SNR shells up to 50 keV.

Non-thermal emission in many sources of interest such as gamma-ray
bursts, AGNs and galaxy clusters originates from synchrotron
radiation of ultra-relativistic particles in turbulent magnetic
fields. The effect of random fields on synchrotron spectra and
time evolution should be accounted for in their models.
Relativistic flows in GRBs, PWNs and AGNs likely produce random
magnetic fields with large amplitudes that should increase the
efficiency of synchrotron radiation from electrons/positrons in
the spectral cut-off regime.Synchrotron radiation in highly
stochastic fields also results in energy band specific variability
of the observed emission and its polarization. These specific
characteristics of the synchrotron images and spectra in the
cut-off regime can be used to study the turbulent magnetic fields
in astrophysical sources.

\acknowledgments
 We thank a referee for constructive comments. A.~M.~B. and
Yu.~A.~U. acknowledge support from RBRF grant 06-02-16884 and
D.~C.~E. from NASA grants NNH04Zss001N-LTSA and 06-ATP06-21.

\bibliographystyle{apj}
\bibliography{synchr}

\clearpage

\end{document}